\documentclass[aip,amsmath,amssymb,reprint]{revtex4-1}
\usepackage{graphicx}
\usepackage{dcolumn}
\usepackage{bm}

\usepackage[utf8]{inputenc}
\usepackage[T1]{fontenc}
\usepackage{mathptmx}
\usepackage{etoolbox}

\makeatletter
\def\@email#1#2{%
 \endgroup
 \patchcmd{\titleblock@produce}
  {\frontmatter@RRAPformat}
  {\frontmatter@RRAPformat{\produce@RRAP{*#1\href{mailto:#2}{#2}}}\frontmatter@RRAPformat}
  {}{}
}%
\makeatother

\begin{document}

\title{Computational Demonstrations of Density Wave of Cooper Pairs and Paired-Electron Liquid in the Quarter-Filled Band - a Brief Review}
\author{Sumit Mazumdar}
\affiliation{Department of Physics, University of Arizona Tucson, AZ 85721}
\email{sumit@physics.arizona.edu}
\author{R. Torsten Clay}
\affiliation{Department of Physics and Astronomy and HPC$^2$ Center for
  Computational Sciences, Mississippi State University, Mississippi State MS 39762}
\date{\today}

\begin{abstract}
There has been strong interest recently in the so-called Cooper pair
density wave, subsequent to the proposition that such a state occurs
in the hole-doped cuprate superconductors. As of now there is no
convincing demonstration of such a state in the cuprate theoretical
literature.  We present here a brief but complete review of our
theoretical and computational work on the paired-electron crystal
(PEC), which has been also experimentally seen in the insulating phase
proximate to superconductivity (SC) in organic charge-transfer solid
(CTS) superconductors. Within our theory, SC in the CTS does indeed
evolve from the PEC. A crucial requirement for the finding of the PEC
is that the proper carrier density of one charge carrier per two sites
is taken into consideration at the outset. Following the discussion of
CTS superconductors, we briefly discuss how the theory can be extended
to understand the phase diagram of the cuprate superconductors that
has remained mysterious after nearly four decades of the discovery of
SC in this family.
\end{abstract}
\maketitle

\vskip 0.25in
{\bf Despite more than 30 years of research, the mechanism of
  superconductivity (SC) in the high-T$_c$ cuprates and other
  unconventional superconductors is still not understood.  In a conventional
  superconductor the superconducting state evolves from a
  metal. However, in the cuprates and many other unconventional
  superconductors, unusual insulating phases are instead found
  adjacent to SC. In this paper we review theoretical work on one of
  these insulating phases, the Paired Electron Crystal (PEC). The PEC
  has been found experimentally in crystalline molecular
  superconductors known as organic charge-transfer solids. We describe
  theoretical work showing how SC can evolve from the PEC, and further
  how the PEC concept can be extended to understand the phase diagram
  of the high T$_c$ cuprates.}
\vskip 0.25in

\section{Introduction}

The theory of charge-density waves (CDWs), driven by either the
Peierls intersite \cite{Su80a} or by the Holstein intrasite
\cite{Holstein59a} electron-phonon (e-p) interactions in molecular
solids are both quite old and well understood by now. The same is true
with the concept of the Wigner crystal (WC), driven by intersite
electron-electron (e-e) interactions \cite{Hubbard78a}.  In all such
cases the electronic wavefunction in the ground state is characterized
by periodic modulation of single electron densities.  Should SC occur
in such systems, the understanding has been that the driving force is
once again the e-p interaction as in the traditional BCS theory
\cite{Bardeen57a}.  Over the past several decades a different
phenomenology, characteristic of many if not most unconventional
superconductors, has attracted significant attention. This is the
appearance of SC not merely proximate to CDW or spin-density wave
(SDW)/antiferromagnetism (AFM) in the phase diagram, but the
observation of actual ``positive correlation'' between spatial broken
symmetry and SC, in that the SC may be evolving from the spatial
broken symmetry state.

Evolution of SC from CDW broken state is in principle likely if
the CDW is ``paired'', in that the electronic
wavefunction exhibits modulation of pairs of charge carriers coupled
into spin-singlets, as opposed to single electrons or holes. We have
referred to this state as the PEC \cite{Li10a,Dayal11a}, a terminology
that existed prior to our work \cite{Moulopoulos92a}. Conceptually it
is possible to visualize SC emerging from the PEC with increasing
lattice frustration or doping that leaves the paired singlets intact
even as they become mobile. While this idea does not necessarily
solve the problem of correlated-electron SC (which requires explicit
demonstration of the transition from the PEC to the state with
long-range superconducting pair-pair correlations), it does give an
understandable starting point that has the added benefit of being
``visual''. 

We present here a brief review of our computational work, and 
show how the results can be used to understand spatial broken symmetries 
(especially CDW correlations) in organic CTS that are structurally related to
CTS superconductors, or in the superconductors themselves in the normal
state proximate to SC. We further 
discuss how the same concepts probably can be extended to understand
hole as well as electron-doped cuprates. It is relevant in this
context to note that the experimentally observed charge ordering in
cuprates has been claimed to be a density wave of Cooper pairs by some
research groups \cite{Cai16a,Hamidian16a,Mesaros16a}. 
Computational results performed by our groups have demonstrated
explicitly the occurrences of the PEC within a correlated-electron
Hamiltonian on various lattices. The essential requirement is commensurability, which in turn occurs at
strictly one particular carrier density. We will also discuss the possibility that moving away from strict commensurability
can lead to SC, or at least a paired-electron {\it liquid} (PEL).

\section{Theoretical Model}

The theoretical model we will consider is the same for all
dimensionalities, the extended Hubbard Hamiltonian with nonzero
intersite Coulomb repulsion and intersite Peierls and intrasite
Holstein e-p couplings,

\begin{eqnarray}
H=-\sum_{\nu,\langle ij\rangle_\nu}t_\nu(1+\alpha_\nu\Delta_{ij})B_{ij} 
+\frac{1}{2}\sum_{\nu,\langle ij\rangle_\nu} K^\nu_\alpha \Delta_{ij}^2 \nonumber \\
+g \sum_i v_i n_i + \frac{K_g}{2} \sum_i v_i^2  
+ U\sum_i n_{i\uparrow}n_{i\downarrow} + 
\frac{1}{2}\sum_{\langle ij\rangle}V_{ij} n_i n_j
 \label{ham}
\end{eqnarray}
In Eq.~\ref{ham}, $\nu$ runs over multiple lattice directions $a$, $b$
and $c$. $B_{ij}=\sum_\sigma(c^\dagger_{i\sigma}c_{j\sigma}+H.c.)$,
$\alpha_\nu$ is the intersite e-p coupling, $K^\nu_\alpha$ is the
corresponding spring constant, and $\Delta_{ij}$ is the distortion of
the $i$--$j$ bond, to be determined self-consistently; $v_i$ is the
intrasite phonon coordinate and $g$ is the intrasite e-p coupling with
the corresponding spring constant $K_g$. The hoppings $t_\nu$ and
intersite Coulomb interaction $V_{ij}$ are between nearest neighbors
(n.n).  We consider only $\rho=\frac{1}{2}$, where there is strongest
likelihood of commensurability-driven density wave composed of
electron pairs. Additionally, much of our original motivation behind
this research came from trying to reach a consistent theory of SC in
organic CTS, which have uniformly carrier density per molecule of
$\rho=\frac{1}{2}$.
 
\section{PEC in one dimension}

\subsection{Experimental Summary}

Our involvement in this research area began with our attempt to
understand the 4k$_F$ metal-insulator (M-I) and 2k$_F$ insulator (I-I)
transitions in the organic CTS MEM(TCNQ)$_2$, which at the time was
understood the most experimentally. Each MEM donor molecule donates
one electron to two TCNQ acceptor molecules, such that the
one-dimensional (1D) stack of TCNQ anions has a carrier density
$\rho=\frac{1}{2}$ and is conducting at high temperatures in spite of
large Hubbard intramolecular repulsion. Below 335 K the system
undergoes M-I transition that is driven by intrastack bond
dimerization \cite{Visser83a,Huizinga79a}. At still lower temperature
of 17.2 K there occur both bond and intramolecular site charge
tetramerization \cite{Visser83a,Huizinga79a} and a gap in the magnetic
spectrum is observed from measurements of magnetic susceptibility
\cite{Huizinga79a}.  Very similar coexisting bond and charge
modulations were seen also in the low temperature tetramerized phase
of TEA(TCNQ)$_2$, which is insulating and dimerized already at high T
\cite{Kobayashi70a,Filhol84a}.  Coexisting bond and charge-modulations
had also been found in the so-called spin-Peierls (SP) phase of the
cationic CT compounds (TMTTF)$_2$PF$_6$ and (TMTTF)$_2$AsF$_6$, in
which the charge on the TMTTF cation is $\frac{1}{2}$.  The {\it
  coexisting} bond and charge tetramerization was not understood at
the time.

It is important to understand in this context precisely {\it why} this
coexistence was difficult to understand within what then was the
traditional approach. Within the strongly correlated scenario the M-I
transition in $\rho=\frac{1}{2}$ is due either to 4k$_F$ bond
dimerization (alternate short and long bonds) or to 4k$_F$ charge
dimerization that results in the WC with alternate charge-rich and
charge-poor sites.  Within existing theories at the time, each dimer
in the bond-dimerized phase was considered as a single ``site'', the
system was considered as an {\it effective} half-filled band
(effective $\rho=1$), and bond tetramerization was the dimerization
within this effective half-filled band.  Charge and bond order
modulations are not expected to coexist within the half-filled band
\cite{Lin00a}.  Starting from the 4k$_F$ WC with alternate sites
occupied by charge carrier, conversely, further bond dimerization
would require unusually strong e-p couplings that could modulate the
distances between ``occupied'' sites two lattice constants apart,
making the bond tetramerization highly unlikely. Furthermore, the
expected pattern of the bond distortion in this case,
strong-strong-weak-weak, has not been observed in any CTS.  The
experimentally observed bond distortion pattern is
strong-weak-strong-weak$^{\prime}$, with weak and weak$^{\prime}$ bond
strengths different.  In what follows we will find that precise
understanding requires discarding the effective half-filled band
concept at the outset.

\subsection{Theory and Computational results.}

As has been discussed extensively elsewhere \cite{Clay19a}, physical
insight to unconditional broken symmetries within Eq.~\ref{ham} can be
obtained by considering the limit of zero e-p coupling. In this limit
and for $U \to \infty$ the Hamiltonian in 1D reduces to the spinless
Fermion Hamiltonian,
\begin{equation}
H_{\rm eff} = V\sum_i n_in_{i+1} - t\sum_i(a^\dagger_ia_{i+1}+a^\dagger_{i+1}a_i)
\label{spinless}
\end{equation}
where $a^\dagger_i$ create spinless Fermions and $n_i=0,1$ only. For
large $V$, the ground state now is the 4k$_{\rm F}$ charge-ordered
phase (CO) $\cdots$1010$\cdots$, where the numbers 1 and 0 refer to
actual site occupancies of 0.5+$\delta$ and 0.5-$\delta$,
respectively. It is essential here to appreciate that the WC is
obtained only for $V = V_c \geq 2|t|$ \cite{Mila93a}.  More
importantly in the present case, extensive numerical calculations by
multiple groups have that V$_c$ {\it increases} as $U$ becomes finite
(for e.g., $V_c$ approaches $3|t|$ at $U=8|t|$)
\cite{Penc94a,Shibata01a,Clay03a} The pertinent question is then what
spatial broken symmetry occurs for $V < V_c$.
\begin{figure}
  \centerline{\resizebox{3.0in}{!}{\includegraphics{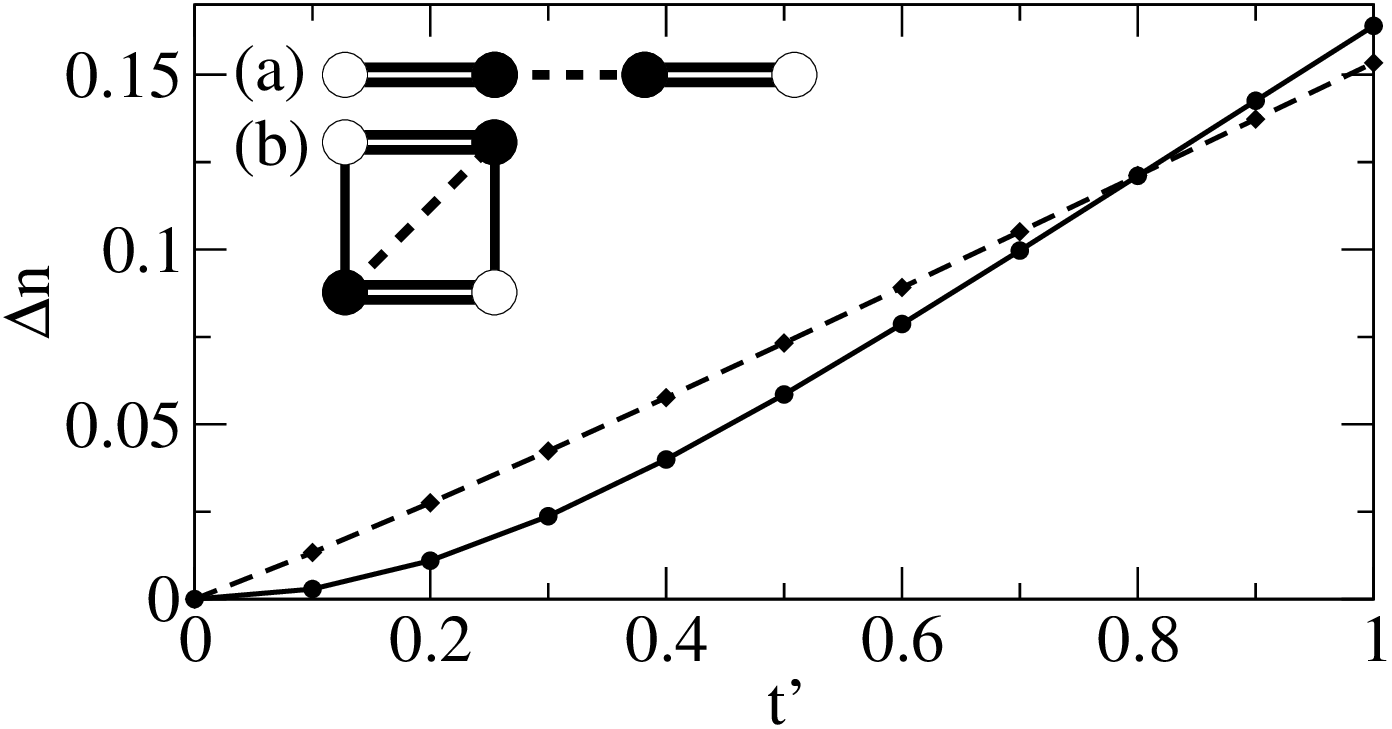}}}
    \caption{Charge difference $\Delta n$ versus hopping integral
      $t^\prime$ for two four-site systems.  Solid lines are for the
      linear four-site system shown in inset (a); dashed lines are for
      the square plaquette in inset (b).  Coulomb interactions are
      $U=4$ and $V=0$.  Double, single, and dashed lines represent
      hopping integrals of strengths $t_1$=1.5, $t_2$=0.5, and
      $t^\prime$, respectively.  In the insets, filled and empty
      circles correspond to charge densities greater or less than
      0.5. Reproduced from Ref. \onlinecite{Dayal11a}. Copyright 2011, American Physical Society.}
  \label{figdimers}
\end{figure}

A conceptual framework that allows a pictorial resolution of this
question is as follows. Consider a single dimer of two sites with one
electron. The electron populations per site are 0.5 each, but the
quantum mechanical wave function for the system is the superposition
$2^{-1/2}[10 + 01]$, where 1 and 0 are site charge densities. If one
now brings two of these dimers together, as in inset (a) of
Fig.~\ref{figdimers}, the composite wave function of the two-dimer
system can be written as $\frac{1}{2}[1010 + 1001 + 0110 + 0101]$. If
the two electrons are in a spin-singlet state then within the simple
Hubbard Hamiltonian the configuration 0110, in which singlet
stabilization can occur from a single n.n. hop that creates a virtual
double occupancy, must dominate over the configurations 1010 and 1001,
in which singlet stabilization requires two and three hops,
respectively. Thus as the singlet bond between the dimers gets
stronger we expect increasing charge difference $\delta$ between sites
belonging to the same dimer (between sites 1 and 2, or between sites 3
and 4 in the linear chain of Fig.~\ref{figdimers}). Our proposed
picture demands that similar charge disproportionation occurs between
members of the same dimer even in the case of the periodic molecule
shown in the inset (b) of Fig.~\ref{figdimers}. In this case the
charges on the sites connected by the diagonal bond must be larger
than 0.5, while the charges on the two other sites must be smaller.

\begin{figure}
  \centerline{\raisebox{1.2in}{(a)}\hspace{0.1in}\resizebox{2.0in}{!}{\includegraphics{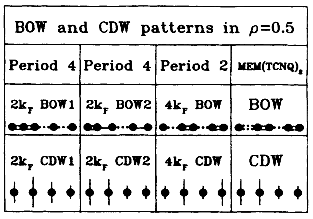}}}
  \centerline{\resizebox{3.0in}{!}{\includegraphics{fig2b}}}
  
  \caption{(a) Schematic of the possible charge and bond distortions
    in a 1D chain with density 0.5. Single, double, and dotted lines
    represent undistorted, short, and long bonds, respectively. The
    length of the vertical lines corresponds to the charge density on
    each site. The last column shows the bond and charge distortions
    found in MEM(TCNQ)$_2$ at low temperature. Reproduced from 
Reference \onlinecite{Ung94a}. Copyright 1994 American Physical Society.
  Zero temperature phase diagram of Eq.~1 in the limit of 0$^+$
    electron-phonon interactions.  
BCDW2 is the 2k$_F$
    BOW2 in (a); BCDW1 is the BOW shown in the 4th column of (a). The
    2k$_F$ CDW1 of (a) only exists at very small $U$ and is not shown
    here.  The $V=\frac{U}{2}$ line indicates the region of physical
    relevance for the organic CTS. Reproduced from Reference \onlinecite{Clay17a}. Copyright 2017 American Physical Society.}
  \label{fig1d}
\end{figure}

The conclusions drawn from the simple 4-atom calculations reported in
Fig.~\ref{figdimers} were confirmed from exact finite size
calculations on N = 8, 12 and 16 finite periodic rings \cite{Ung94a},
and quantum Monte Carlo calculations\cite{Clay17a} of up to N = 96.
In all cases bond-order wave (BOW) and 2k$_F$ site-diagonal
charge-density wave (CDW) coexist provided $V < V_c$.  The results are
summarized in Fig.~\ref{fig1d}.  For noninteracting electrons or very
small $U$ the ground state broken symmetry consists of coexisting
2k$_F$ (period 4) BOW1 and 2k$_F$ CDW1, whereas for larger $U$ the
ground state is a superposition of a 2k$_F$ CDW2 and a BOW with the
pattern strong-medium-weak-medium (SMWM) (2nd column of
Fig.~\ref{fig1d}(a)).  For stronger $U$ and $V$ the coexisting BOW
instead has the pattern strong-weak-strong-weak$^\prime$
(SWSW$^\prime$, 4th column of Fig.~\ref{fig1d}(a)).  The
experimentally determined bond distortions and charge densities in
MEM-(TCNQ)$_2$ \cite{Huizinga79a,Ung94a} correspond exactly to this
phase. The agreement with theory is true \cite{Kobayashi70a,Filhol84a}
also for the lowest temperature phases in TEA(TCNQ)$_2$.  Notice that
the large charge densities in the BCDW1 and BCDW2 phases are coupled
by spin-singlet bonds, and hence the CO phase does correspond to a
density wave of spin singlets.  The agreements are not surprising, and
follow directly from the conceptual framework of Fig.~\ref{figdimers},
that is based only on the principle of charge-spin coupling in
correlated $\rho=\frac{1}{2}$ systems.

\section{The weakly two-dimensional regime}

\subsection{Experimental Summary}

The first organic superconductors, (TMTSF)$_2$X, where X are
monovalent anions PF$_6^{-}$, AsF$_6^{-}$, ClO$_4^{-}$ etc.  were
discovered in the early 1980s \cite{Jerome80a}. Unlike the TMTTF-based
systems which are quasi-one dimensional with spin-Peierls ground
states, their Se analogues should be considered weakly two-dimensional
(see below). The compound (TMTSF)$_2$PF$_6$ has been studied the most
intensively, and is considered the prototype member of this
family. This compound exhibits metallic behavior at high temperature,
but under ambient pressure there occurs transition to an
incommensurate SDW at 12.5 K. The occurrence of the SDW, instead of
the SP state, is direct evidence for non-negligible two-dimensional
hopping between the TMTSF cations. Computations lead to
parametrizations $t_{||}$ = 0.1-0.2 eV, $t_{\perp}$ = 0.01-0.02 eV,
where $t_{||}$ and $t_{\perp}$ are the intra- and interstack hole
hopping integrals between the TMTSF cations.  Application of moderate
pressure (6.5 kbar) leads to an apparent SDW-to-SC transition, with
critical temperature T$_c$ = 1.2 K. Several other TMTSF compounds
exhibit similar SDW-to-SC transitions.  These observations had led to
early applications of the spin-fluctuation theory of SC to this family
of CT solids, as discussed below.

The intrastack intermolecular distances in (TMTSF)$_2$X once again
alternate and exhibit dimerization.  Assumption of equal charge
densities on the individual molecular sites within each dimer then
leads naturally to the effective $\rho=1$ description for (TMTSF)$_2$X
(as in (TMTTF)$_2$X), with the difference that interstack couplings
are now non-negligible.  The ground state is then a 2D commensurate
$\rho=1$ AFM. Within this picture the superconducting transition is
from AFM-to-SC, to be understood within the spin-fluctuation
mechanism.

The above theoretical scenario was shown to be overly simplistic based
on X-ray scattering experiments performed by Pouget and Ravy in 1997
\cite{Pouget97a}.  These authors found that the broken symmetry state
in (TMTSF)$_2$PF$_6$ was a coexisting CDW-SDW state, which should not
occur in true $\rho=1$. Even more importantly, while a coexisting
CDW-SDW state can be expected within the then standard WC models of
strongly correlated $\rho=\frac{1}{2}$, the periodicities of CDW and
SDW in this latter case are expected to be 4k$_F$ (alternate sites
occupied by holes) and 2k$_F$ (opposite spins on n.n.  occupied
sites), respectively. The X-ray scattering measurement indicated {\it
  identical} 2k$_F$ periodicity (period 4) for both the CDW and the
SDW. Pouget and Ravy labeled this broken symmetry ``unprecedented''.
Importantly, transition to a commensurate SDW was also found
\cite{Nagata13a}.  SC evolves from the coupled 2k$_F$ CDW-SDW and its
mechanism has to be understood within this context.

\subsection{Theory and Computational results.}

\begin{figure}
  \centerline{\resizebox{2.0in}{!}{\includegraphics{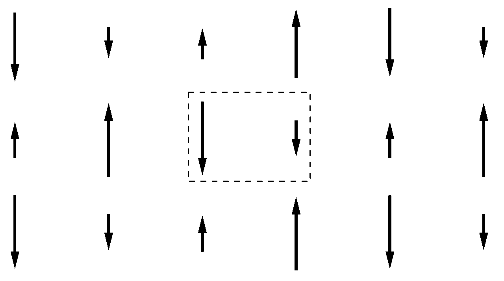}}}
  \caption{The 2k$_F$ period 4 coexisting CDW-SDW found at
    $\rho$=0.5. The length of the arrows corresponds to the charge
    density on each site. The dashed box surrounds one dimer unit.
  Reproduced from Reference \onlinecite{Mazumdar99a}. Copyright 1999 American Physical Society.}
  \label{fig2d}
\end{figure}

The physical arguments above for unequal intradimer charge occupancies
in the 1D limit applies also to the weakly 2D antiferromagnetic case,
as the stabilization of the antiferromagnet (over high spin states)
originates also from virtual CT that creates double occupancies. Thus
neighboring interdimer sites with shorter interdimer separation
(stronger interdimer coupling) must have charge densities larger than
corresponding interdimer site pairs with longer interdimer
separation. This was proved within Eq.~\ref{ham} from both 1D and 2D
numerical calculations \cite{Mazumdar99a}. The 1D calculation involved
adding external 2k$_F$ AFM potential to the static version of
Hamiltonian \ref{ham}, for fixed $U$ and $V$, and determining the
tendency to 2k$_F$ bond-charge distortion as a function of the
strength of the AFM potential. The energy gained upon bond distortion
increases with the strength of the external AFM potential, which is an
indicator {\it co-operative coexistence} of 2k$_F$ BOW-CDW and 2k$_F$
AFM. The 2D calculations were done using the constrained path
renormalization group (CPMC) approach for 4 coupled chains of length
12 sites each, periodic along both directions, with $t_{\perp}
=0.1t_{||}$. Lowest ground state energy was found for $\pi$-phase
difference between BOWs on neighboring chains, which was therefore
adopted. Measurements of charge-charge and spin-spin correlations then
established the coexisting 2k$_F$ CDW-SDW shown in
Fig.~\ref{fig2d}. Once again there is charge disproportionation within
each dimer unit cell, and the CDW is period 4. The coexisting 2k$_F$
SDW results from fractional charges within the dimer unit cell having
the same spin, which is necessarily true as they together constitute a
single hole. As in the 1D limit, the density wave of Fig.~\ref{fig2d}
is also paired, with the pair now being antiferromagnetically coupled
as opposed to being spin singlet. It is then not inconceivable that as
these pairs begin hopping between chains with pressure-induced
increase in $t_{\perp}$ a superconducting state is reached.

\section{The strongly two-dimensional regime}

\subsection{Experimental Summary}

The 2D CTS feature a great variety of compounds and behavior. There
exist multiple recent reviews covering these systems
\cite{Clay19a,Dressel20a,Naito21a}.  Here we limit ourselves to only
those discussions that pertain to the PEC.  As with the weakly 2D
superconducting CTS, these are also mostly 2:1 systems with general
chemical formula A$_2$X, where A is the organic molecular component
and X are inorganic monomer with charge $-1$, so that the average
charge on the organic cation is $+\frac{1}{2}$. There also exist
anionic 1:2 superconductors, with now organic anion charge of
$-\frac{1}{2}$.  We limit ourselves to cationic 2:1 superconductors
only, as the physics of the anionic compounds are very similar and can
be understood within the same theoretical approach \cite{Clay19a}

The largest number of 2D superconducting CTS have A = (BEDT-TTF),
hereafter ET. While there exist other superconducting A$_2$X with
cationic component different from ET, because of the limited scope of
this presentation in what follows we limit ourselves to A = ET only,
as these exhibit the full range of behavior seen in the 2:1 cationic
superconductors. (ET)$_2$X occur in different crystalline forms, which
are labeled as $\kappa$, $\alpha$, $\beta$ and $\theta$. In the
following we give broad summaries of the experimental observations for
each of these crystal structures.

\noindent\underbar{$\kappa$-(ET)$_2$X} These ET compounds with X =
Cu(NCS)$_2$, Cu[N(CN)$_2]$Cl, Cu[N(CN)$_2$]Br, Cu$_2$(CN)$_3$
etc. have been among the most intensively studied CTS
superconductors. The cation layer is characterized by strongly
dimerized anisotropic triangular lattices with one hole per
ET$_2^{1+}$ dimer.  Relatively few have ground states that are AFM or
quantum spin liquid (QSL), but much of the theoretical attention have
centered around these, because of SC being proximate to these magnetic
states, thus providing an apparently tangible connection to the
observation in the cuprates.  The other $\kappa$-phase materials
exhibit CO or spin gap (SG). SG has been observed in some
charge-ordered $\kappa$-(ET)$_2$X. SC is spin-singlet, and the order
parameter has nodes. There is evidence for intra-dimer charge
fluctuations and relatively strong lattice effects.

\noindent\underbar{$\beta$-(ET)$_2$X} These contain strongly
interacting one-dimensional cation stacks, with the molecular planes
nearly perpendicular to the stacking axes. Some of these exhibit
ambient pressure SC, which is often proximate to nonmagnetic
insulating states with pronounced lattice distortion that is often
period-4. Others exhibit transition to SC under pressure. CO-to-SC
transition is common among $\beta$-(ET)$_2$X. In all cases where the
pattern of the CO or bond distortion is known, they appear to be
different from that of the simple WC, in that in more than one
direction the CO pattern is $\cdots 1100 \cdots$. SG often accompanies
the CO  at the lowest temperatures.

\noindent\underbar{$\theta$-(ET)$_2$X} In the $\theta$ structure
molecules in neighboring stacks are tilted with respect to each other
by dihedral angle 100$^o$-140$^o$. MI transition to CO is common to
$\theta$-ET and is often followed by a lower temperature transition to
a nonmagnetic state with SG. The CO consists of a stripe pattern
(``horizontal'') which is different from a simple WC and has charge
occupancies $\cdots 1100 \cdots$ along the two most strongly coupled
directions. SC occurs in X = I$_3$.

\noindent\underbar{$\alpha$-(ET)$_2$X} The $\alpha$ structure is
nearly identical to the $\theta$ structure with the difference that
the periodicity is doubled already in the room-temperature structure
in the stacking direction by a weak dimerization. In addition, the
stacks are often inequivalent. CO is common, with the CO pattern
corresponding to horizontal stripe. CO and SG transitions occur at the
same temperature in X = I$_3$. In some compounds there likely is
coexisting CDW-SDW.

\subsection{Theory and Computational Results}

\begin{figure}
  \centerline{\resizebox{3.0in}{!}{\includegraphics{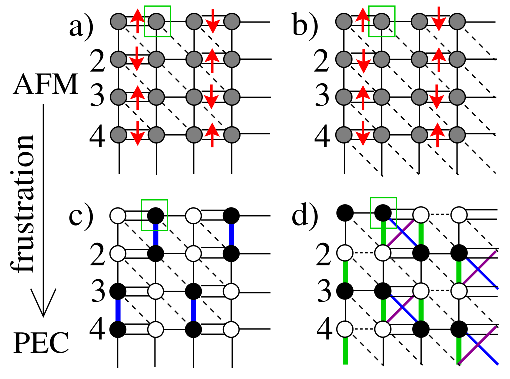}}}
  \caption{Schematic of the AFM and PEC states as seen in a 4$\times$4
    lattice.  The hopping integral boundary conditions
    are periodic (PBC) along $x$ and $y$ directions in both (a) and
    (b) and open (OBC) and PBC in (a) and (b), respectively, along the
    $x-y$ directions.  Double and thick lines represent strong bonds;
    thin lines represent weak bonds. Dashed lines represent
    $t^\prime$, whose strength is varied.  Charge densities as
    indicated by grey circles are uniform in (a) and (b), and spin
    ordering corresponds to AFM. (c) and (d) shows the PEC state for
    $t^\prime > t^\prime_c$. Here black and white circles present
    charge-rich and charge-poor sites. Singlet bonds form between
    charge-rich sites.  There occur periodic arrangement of spin-pairs
    along $y$ and $x-y$ directions in (a), and along $x$ and $x-y$
    directions in in (b). The box marks the reference site for
    spin-spin correlations shown in Fig.~\ref{fig2dpecspin}. Numbers
    are the chain indices used in Fig.~\ref{fig2dpecspin}(a) and (b).
Reproduced from Reference \onlinecite{Li10a}. Copyright 2010 Institute of Physics Publishing.}
  \label{fig2dpec}
\end{figure}

The goal of the theoretical work here was to demonstrate geometrical lattice frustration-driven AFM-to-PEC transition, which we believe 
precedes the transition to SC in systems where apparently AFM and SC are proximate. Calculations were done starting from the 
bond dimerized limit along the $x$-direction, since (a) bond dimerization is essential to obtain AFM in $\rho=\frac{1}{2}$,
and (b) CTS with AFM ground states are universally dimerized (see discussion on AFM in $\kappa$-(ET)$_2$X above). The calculations were done on the 4$\times$4 lattice with both zero e-p interaction and
externally imposed rigid dimerization (see Fig.~\ref{fig2dpec}(a)), as
well as with nonzero e-p interactions along the $x$ and $y$ directions
(see Fig.~\ref{fig2dpec}(b)). In the first case we chose periodic
hoppings along the $x$ and $y$ directions but open boundary condition
(OBC) along the $x-y$ direction. The parameters for the OBC
calculations were $t_x=t \pm \delta_t$, $t_y=t$, $t=1$,
$\delta_t=0.2$, $\alpha_{\nu}=g=0$, with in-phase bond dimerization
between consecutive chains along the $y$-direction that gave
commensurate AFM.  In the second case all hoppings were periodic
(periodic boundary condition, PBC) and e-p interactions were
explicitly included. These latter parameters were $\alpha_x=1.3$,
$\alpha_y=1.0$, $\alpha_{x-y}=0$, $K_{\alpha}^x=K_{\alpha}^y=2$,
$g=0.1$ and $K_g=2$. For the first set of calculations we report below
we chose $U=4$, and $V_x=V_y=1$, $V_{x-y}=0$. The intersite Coulomb
interactions were not large enough at the chosen $U$ to give the WC
ground state (note that nonzero $V_{x-y}$ further destabilizes the
WC). The parameter $t_{x-y}=t^\prime$ is the variable that drives the
AFM-PEC transition.

\begin{figure}
  \centerline{\resizebox{3.2in}{!}{\includegraphics{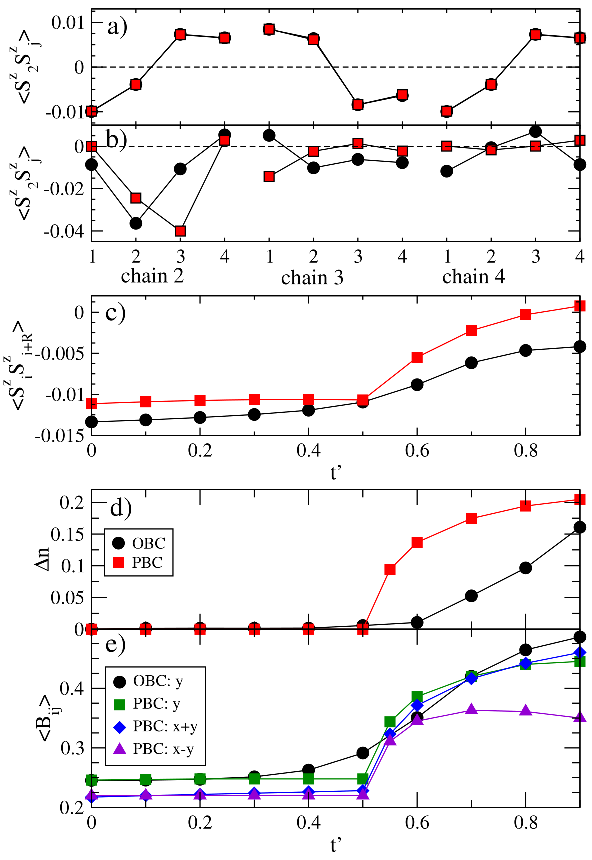}}}
  \caption{(a) Spin-spin correlations $\langle S_i^zS_{j}^z \rangle$
    between site 2 (marked with a box in Fig.~\ref{fig2dpec}) and
    sites 1--4 in chains 2--4 for $t^\prime$=0. In all panels, circles
    and squares correspond to OBC and PBC calculations, respectively.
    (b) Spin-spin correlations as in (a), but with $t^\prime$=0.7.
    (c) Spin-spin correlations between sites of the most distant
    dimers versus $t^\prime$ (d) Charge disproportionation versus
    $t^\prime$ (e) Bond orders between between pairs of n.n. sites
    forming localized spin-singlets. For the OBC these are along the
    $y$ direction (see Fig.~\ref{fig2dpec}(c)). For the PBC lattice
    bonds along the $y$, $x+y$, and $x-y$ all change at the PEC
    transition (Fig.~\ref{fig2dpec}(d)). These are plotted using
    squares, diamonds, and triangles, respectively. Reproduced from Reference \onlinecite{Li10a}. Copyright 2010 Institute of Physics Publishing.}
  \label{fig2dpecspin}
\end{figure}

In Fig.~\ref{fig2dpecspin}(a) we show the z-z spin-spin correlation
functions for $t^\prime = 0$ between a fixed site (marked with box on
each lattice in Fig.~\ref{fig2dpec}) and sites $j$, labeled
sequentially 1, 2, 3, 4 from the left, on neighboring chains labeled
2, 3, 4 in Fig.~\ref{fig2dpec}. In Fig.~\ref{fig2dpecspin}(a) only,
the average spin-spin correlation with each chain has been shifted to
zero (note dotted line for $\langle S_2^zS_j^z \rangle=0$ in
Figs.~\ref{fig2dpecspin}(a)-(b)) in order to clearly show the AFM
pattern, which are $\cdots - - + + \cdots$ and $\cdots + + - - \cdots$
on the nearest and next nearest chains, indicating N\'eel ordering of
the dimer spin moments in both lattices. The loss of this pattern in
Fig.~\ref{fig2dpecspin}(b) for large $t^\prime = 0.7$ indicates loss
of AFM order. In Fig.~\ref{fig2dpecspin}(c) we plot the spin-spin
correlation between maximally separated dimers, which measures the
strength of the AFM moment.  This correlation is nearly constant until
$t^\prime \sim 0.5$, beyond which the AFM order is destroyed.

Fig.~\ref{fig2dpecspin}(d) shows the difference in charge densities
$\Delta n$ as a function of $t^\prime$. There is a rapid increase in
$\Delta n$, starting from zero, for $t^\prime > t^\prime_c$ with both
lattices. Simultaneously with CO, there is a jump in the bond orders
$\langle B_{ij} \rangle$ between the sites that form the localized
spin-singlets. This is shown in Fig.~\ref{fig2dpecspin}(e). These bond
orders are by far the strongest in both lattices for $t^\prime >
t^\prime_c$. The spin-spin correlation between the same pairs of sites
becomes strongly negative at the same $t^\prime$, even as all other
spin-spin correlations approach zero (Fig.~\ref{fig2dpecspin}(b)),
indicating spin-singlet character of the strongest bonds. The
simultaneous jumps in $\langle S_i^zS_{i+R}^z \rangle$, $\Delta n$ and
$\langle B_{ij} \rangle$ at the same $t^\prime$ give conclusive
evidence for the AFM-PEC transition shown in Fig.~\ref{fig2dpec}.

\begin{figure}
  \centerline{\resizebox{3.0in}{!}{\includegraphics{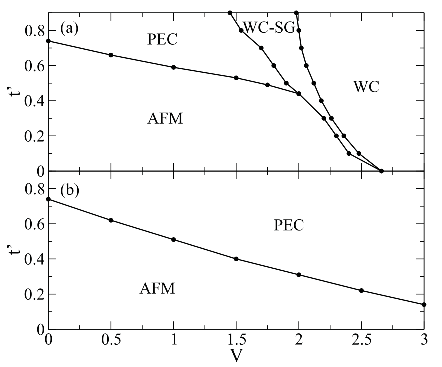}}}
  \caption{(a) Phase diagram for the 4$\times$4 lattice of
    Fig.~\ref{fig2dpec}(b) as a function of $t^\prime$ and
    $V=V_x=V_y$, $V^\prime$=0, with $U=6$, $\alpha_\nu$=1.1, $g$=0.1,
    and $K^\nu_\alpha=K_g$=2.0. (b) Same as (a), but with
    $V=V_x=V_y=V^\prime$. 
Reproduced from Reference \onlinecite{Dayal11a}. Copyright 2011 American Physical Society.}
  \label{figpecphase}
\end{figure}

Fig.~\ref{figpecphase} shows how e-e interactions $U$ and $V$ affect
the PEC \cite{Dayal11a}. As shown in Fig.~\ref{figpecphase}(a),
increasing $U$ moderately increases $t^\prime_c$. The figure also
shows the effect of the n.n.  Coulomb interaction $V$.  Here the phase
diagram depends critically on the form assumed for ${\langle V_{ij}
  \rangle}$. With $V_x = V_y = V$ but $V^\prime = 0$, the WC is found
for sufficiently strong $V$, along with a narrow region where it
coexists with a SG. The very narrow width of the WC-SG phase in the
figure indicates the small likelihood of the WC coexisting with SG in
real materials.  Equally importantly, most CTS lattices are partially
triangular, and the assumption that $V^\prime = 0$ is not
realistic. For $V_x = V_y = V^\prime$ the WC and the PEC have the same
classical energies; as shown in Fig.~\ref{figpecphase}(b) in this case
the WC is completely replaced by the PEC.

\section{PEC and Superconducting Correlations}

To summarize the previous sections, the PEC is computationally arrived
at $\rho \simeq \frac{1}{2}$ independent of dimensionality, with
geometric lattice frustration playing a key co-operative
role. Experimentally, there is evidence for the PEC in many different
families of CTS with 2:1 cation:anion composition ratio and in the few
known 1:2 compounds that exhibit SC \cite{Clay19a}. This is strong
indication that SC evolves from the PEC following weak doping or
increased frustration that lead to weak deviation from lattice
commensurability. In the following we briefly review numerical results
that support this viewpoint \cite{Gomes16a}.

Our calculations were within the static extended Hubbard Hamiltonian without e-p interactions,
\begin{equation}
  H = -\sum_{\langle ij\rangle,\sigma}t_{ij} B_{i,j,\sigma} + U\sum_i n_{i,\uparrow}
  n_{i,\downarrow} + \frac{1}{2}\sum_{\langle i j\rangle}V_{ij} n_i n_j
  \label{hamuv}
\end{equation}

where all terms have the same meanings as in Eq.~(1). Our calculations
are for anisotropic triangular lattices with ${t_{ij} = t_x, t_y,
    t_{x+y}}$, periodic in all three directions.  We express all
quantities with dimensions of energy in units of $t_x$ ($t_x$ =
1). The bulk of our calculations are for $t_y = 0.9$ and $t_{x+y}$
only slightly smaller than $t_y$. AFM or CO dominate at weaker
frustrations, as seen in the previous sections.  An essential
condition for obtaining numerically precise results within the
approximate quantum mechanical approaches used by us is having
nondegenerate occupancies of single-particle levels in the
noninteracting limit.  The small inequalities between the hopping
integrals maximizes the number of lattice sites $N$ for which the
ground state at or near quarter-filling is nondegenerate.  used by us.
With this constraint of nondegenerate ground states, and consideration
of lattices for which $L_y \geq L_x/2$, our only possible choices at
the time these computations were done were 10 $\times$ 10, 10 $\times$
6, and 6 $\times$ 6. We considered the 4 $\times$ 4 lattice in
addition, which in spite of degeneracy at $\rho = 0.5$ can be treated
exactly.

We define the standard singlet pair-creation operators,
\begin{equation}
  \Delta^\dagger_i = \sum_\nu g(\nu)\frac{1}{\sqrt{2}}(c^\dagger_{i,\uparrow}c^\dagger_{i+\vec{r_\nu},\downarrow}
  - c^\dagger_{i,\downarrow}c^\dagger_{i+\vec{r_\nu},\uparrow}),
\end{equation}
For $d_{x^2-y^2}$ symmetry, $g(\nu) = {1,-1,1,-1}$ for $\vec{r}_{\nu}
= {\hat{x},\hat{y},-\hat{x},-\hat{y}}$, respectively. For $d_{xy}$
symmetry, $g(\nu) = {1,-1,1,-1}$ for $\vec{r}_{\nu} =
{\hat{x}+\hat{y}, -\hat{x}+\hat{y}, -\hat{x}-\hat{y},
  \hat{x}-\hat{y}}$, respectively. We calculated the
distance-dependent pair-pair correlations $P(r)$ ($r\equiv|\vec{r_i}
-\vec{r_j}|$) and show here the average long-range pair-pair
correlation $\bar{P} = N_P^{-1} \sum_{|\vec{r}|>2} P(r)$, where $N_P$
is the number of terms in the sum \cite{Huang01a}.

\begin{figure}
  \centerline{\resizebox{3.3in}{!}{\includegraphics{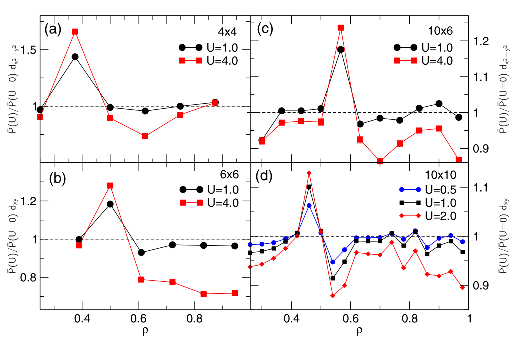}}}
  \caption{Average long-range pair-pair correlation $\bar{P}(U)$
    normalized by its uncorrelated value for (a) 4$\times$4,
    (b) 6$\times$6, (c) 10$\times$6, and (d) 10$\times$10 anisotropic
    triangular lattices, for $t_y$=0.9 and $t_{x+y}$=0.8. 4$\times$4
    results are exact; 6$\times$6 and 10$\times$6 results used the
    PIRG method; and 10$\times$10 the CPMC method. 
Reproduced from Reference \onlinecite{Gomes16a}. Copyright 2016 American Physical Society.}
  \label{figsc1}
\end{figure}

We found $d_{x^2-y^2}$ and $d_{xy}$ symmetries to dominate over
$s$-wave symmetries. Further, for each lattice only one of the two
d-wave channels is relevant; $d_{x^2-y^2}$ for 4 $\times$ 4 and 10
$\times$ 6, and $d_{xy}$ for 6 $\times$ 6 and 10 $\times$ 10. Note
that the distinction between $d_{x^2-y^2}$ and $d_{xy}$ symmetries is
largely semantic in the strongly frustrated regime we investigate.
The complete results are summarized in Fig.~\ref{figsc1}. For each
lattice $\bar{P}(U)/\bar{P}(U=0)>1$ for a single $\rho$ that is either
exactly 0.5 or one of two closest carrier fillings with closed-shell
Fermi-level occupancy at $U = 0$. Pair correlations are {\it
  suppressed} by $U$ at all other $\rho$, including the region
$0.7<\rho<1$ that has been extensively investigated in recent years.
The unique behavior of $\bar{P}(U)/\bar{P}(U=0)$ at or near $\rho=0.5$
cannot be merely coincidences, in view of what we have discussed in
the previous sections.

\begin{figure}
  \centerline{\resizebox{3.0in}{!}{\includegraphics{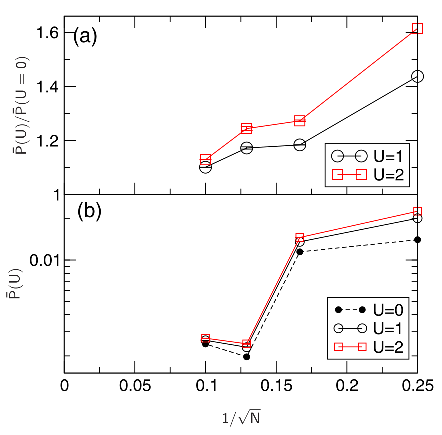}}}
  \caption{Finite-size scaling of the $\rho\approx$ 0.5 peak of
    (a) $\bar{P}(U)/\bar{P}(U=0)$ and (b) $\bar{P}(U)$. $N$ is the
    total number of lattice sites.
    Reproduced from Reference \onlinecite{Gomes16a}. Copyright 2016 American Physical Society.}
  \label{figsc2}
\end{figure}

Rigorous finite-size scaling of the pair-pair correlations is
difficult, both because the enhancement of the pair-pair correlations
occur for different symmetries ($d_{x^2-y^2}$ versus $d_{xy}$) as well
as at slightly different densities for the different lattices. In
Fig.~\ref{figsc2} we have shown our attempt to finite-size scaling of
$\bar{P}$ at relatively small $U$. Long-range superconducting
correlations would require that $\bar{P}$ to extrapolate to finite
value as $N \to \infty$.  This is clearly not the case as seen in
Fig.~\ref{figsc2}(a). We point out, however, that if we ignore
the data points for the largest lattice, 10 $\times 10$,
$\bar{P}(U)/\bar{P}(U=0)$ does converge to a nonzero value, $\sim
0.2$. 

There are multiple possible interpretations of the apparent absence of
long-range superconducting correlations. One possibility is that the
state we are finding is a PEL that is asymptotically close to a
superconducting state with long-range order.  It is conceivable that
phase coherence and long-range pairing is is reached only upon
inclusion of e-p interaction, but there should be no doubt that the
state evolves from a PEC as commensurability effects are reduced.

Calculations similar to the above were done also for the $\kappa$-CTS
lattice geometry \cite{DeSilva16a}, with parameters corresponding to
$\kappa$-(ET)$_2$Cu[N(CN)$_2$]Cl and $\kappa$-(ET)$_2$Cu$_2$(CN)$_3$
(hereafter $\kappa$-Cl and $\kappa$-CN, respectively), which have AFM
and valence bond solid (VBS) ground states, respectively. The
calculations were done for triangular dimerized lattices with 32 and
64 sites (where each site is an ET cation), with intra- and interdimer
separations between sites and hopping integrals taken from the
literature. Although the electron-concentration $\rho_e$ in the ET
cation in these compounds is fixed at 1.5, our calculations were for
the full range of concentration $1 < \rho_e < 2$. We used the Path
Integral Renormalization Group (PIRG) approach for all $\rho_e$ for 32
sites and for $\rho_e=1.5$ for 64 sites. For the other densities on 64
sites we used the CPMC approach.  We calculated both the spin-spin
structure factor and superconducting pair-pair correlations with four
different pairing symmetries, each close to being $d$-wave.

The calculated spin-spin structure factors at $\rho_e=1.5$ correctly
predicted that $\kappa$-CN was further away from AFM than $\kappa$-Cl,
but true long range AFM order was absent even for the latter in our
calculations. We interpreted this to be in agreement with the overall
behavior of the $\kappa$ family, within which relatively few have AFM
ground states, with some members even exhibiting CO.  The
computational results, taken together with the experimental
observations, are in broad agreement with our demonstration in
Figs.~\ref{fig2dpecspin} of the proximity in phase space between AFM
and PEC. Our computational results of the superconducting pair-pair
correlations were very similar to those in Fig.~\ref{figsc1} for the
triangular lattices: for both the $\kappa$-Cl and $\kappa$-CN
lattices, and for both 32 and 64 sites the pair-pair correlations were
enhanced by Hubbard $U$ only for $\rho_e$ exactly 1.5 or immediately
proximate to this, for the same pairing symmetry \cite{DeSilva16a}.

\section{Possible application of the PEC theory to cuprate superconductors}

\begin{figure}
  \centerline{\resizebox{3.3in}{!}{\includegraphics{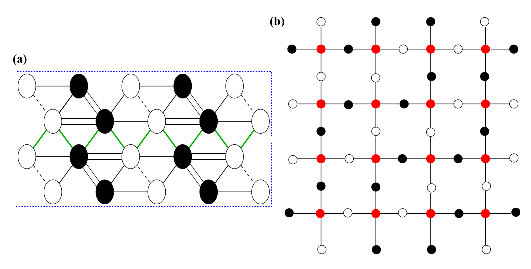}}}
  \caption{(a) PEC charge order in the 2D $\rho=\frac{1}{2}$
    anisotropic triangular lattice Filled and unfilled circles
    correspond to charge-rich and charge-poor sites; the strongest
    bonds are represented by double lines and the weakest by dotted
    lines. Filled circles connect by double lines are spin-singlet
    paired. Reproduced from Reference \onlinecite{Gomes16a}. Copyright 2016 American Physical Society.
(b) Paired charge order in the checkerboard
    $\frac{1}{4}$-filled oxygen sublattice of the CuO$_2$ plane
    following valence transition.  The cations are uniformly
    Cu$^{1+}$. Migration of holes from the erstwhile Cu$^{2+}$ ions
    generates the $\rho \simeq \frac{1}{2}$ O-lattice. Filled and
    unfilled circles correspond to O$^{1-}$ and O$^{2-}$ ions,
    respectively. Pairs of O$^{1-}$ ions linked by the same
    closed-shell Cu$^{1+}$ will be spin-paired. The figure corresponds
    intertwined density waves with periodicities (2$\pi/4a_0$, 0) and
    (0, 2$\pi/4a_0$). Reproduced from Reference \onlinecite{Mazumdar18a}. Copyright 2018 American Physical Society.}
  \label{figcuprate}
\end{figure}

Fully 36 years after the discovery of SC in the cuprates there is
little progress towards understanding of this and related phenomena in
these materials.  Theoretical and computational investigations of
cuprates have largely been within the single-band copper(Cu)-only
Hubbard Hamiltonian, based on (a) the occurrence of commensurate AFM
in the undoped semiconductors, and (b) the Zhang-Rice theory which
claimed that the three-band description of the doped cuprates that
does retain the oxygen (O)-ions can be reduced to the Cu-only one-band
model \cite{Zhang88a}. This approach is encountering serious impasse.
First, recent careful computational studies by many different groups
have found absence of SC within the doped 2D Hubbard model with n.n.
only hopping for carrier concentration 0.7 - 0.9
\cite{Zhang97b,Qin20a,Vaezi21a}, as well as with more distant hoppings
for the hole-doped case \cite{Jiang21a,Jiang22b}.  Our earlier results
in Fig.~\ref{figsc1} had found the same absence of SC in this carrier
concentration range. As of writing this problem has not been resolved.
More importantly, the original Zhang-Rice work had not included the
direct O-O hopping that characterizes real materials
\cite{Hirayama18a}. We have recently demonstrated the absence of quasi
long-range superconducting correlations within the three-band
Hamiltonian in the two-leg Cu$_2$O$_3$ ladder \cite{Song23a}, in
direct contradiction to theoretical results obtained within the
one-band Hamiltonian for the two-leg ladder
\cite{Noack94a,Noack97a}. The strong inequivalence between one-band
and three-band ladder results indicates the breakdown of the
Zhang-Rice theory upon inclusion of direct O-O hopping
\cite{Song23a}. There is no reason to believe that the validity of the
theory will be restored upon going from the two-leg ladder to the 2D
layer.

Beyond the above, satisfactory explanations of the spatial broken
symmetries within the existing theoretical approaches have still not
been reached.  Experimental observations that continue to pose severe
challenges include: (a) density wave of Cooper pairs
\cite{Cai16a,Hamidian16a,Mesaros16a} that we have already mentioned;
(b) momentum space CO periodicity $((Q,0);(0,Q))$, which is often
interpreted as a quasi-1D stripe but in reality describes {\it
  intertwined orthogonal stripes}; (c) saturation with doping of Q to
$2\pi/4a_0$, where $a_0$ is the lattice constant, and finally, (d)
simultaneous breaking of translational and C$_4$ rotational
symmetries, with the broken symmetry state characterized by
inequivalency of O-ions belonging to the same CuO$_2$ unit cell
\cite{Mukhopadhyay19a}. Each of these observations lie outside the
scope of not merely the one-band Hubbard model but even the simple
three-band model Hamiltonian.  Theoretical and experimental
developments, taken together, suggest that the fundamental assumptions
that have gone into understanding SC in cuprates needs serious
reexamination.

Simultaneous resolution to all of the above issues is reached within
the dopant-induced valence-transition hypothesis that we have proposed
for cuprates \cite{Mazumdar18a,Song23b}, within which the PEC occurs
naturally following the transition. The hypothesis is based on three
separate but closely-related observations in transition metal oxides
and heavy fermions that go beyond the concepts that have been used to
understand cuprates until now.  First, recent years have seen
discoveries of {\it negative charge-transfer gap} in a large number of
oxides and chalcogenides in which the true charge on the metal cation
is M$^{(n-1)+}$ with consequent {\it noninteger charge} on O-ions,
instead of simple M$^{n+}$ and O$^{2-}$ as would be expected from the
chemical formula. A partial list includes: (i) BaBiO$_3$, with Bi-ion
charge uniformly 3+ (instead of alternating 3+ and 5+, as was believed
until recently), (ii) rare-earth (RE) nickelates (RE)NiO$_3$ with
Ni$^{2+}$ (instead of Ni$^{3+}$), (iii) FeO$_2$ with Fe$^{3+}$
(instead of Fe$^{4+}$), (iv) CrO$_2$ with Cr$^{3+}$ (instead of
Cr$^{4+}$), (v) AuTe$_2$ with Au$^{1+}$. etc. (see \cite{Song23b} for
extended discussion and original citations). Second, as noted by us,
in each of these cases M$^{(n-1)+}$ electron occupancy is exactly
closed-shell (d$^{10}$, t$_{2g}^6$) or exactly $\frac{1}{2}$-filled
(d$^5$, t$_{2g}^3$, e$_g^2$) for the specific crystal structure,
indicating that the M$^{(n-1)+} \to$ M$^{n+}$ ionization energy
(I. E.)  is higher than usual, a necessary condition for the system to
have negative charge-transfer gap. This is where we further note that
that the electron occupancy d$^{10}$ of Cu$^{1+}$ necessarily implies
that the second I. E. of Cu (Cu$^{1+}$ $\to$ Cu$^{2+}$) is unusually
high relative to the second I. E. of other 3d-elements
\cite{Mazumdar18a}, indicating that cuprates are naturally very close
to the boundary between positive and negative charge transfers.  The
third observation is that pressure- and temperature-driven transitions
between different charge configurations that are close to the same
boundary have been known in organic donor-acceptor charge-transfer
complexes and heavy fermion compounds for over four decades (see
\cite{Song23b} for complete discussions and citations). We therefore
hypothesize that doping drives a similar valence transition in the
layered cuprates, due to the reduced Madelung energy gain and greater
O-hole bandwidth upon doping \cite{Song23b}.

The valence-transition hypothesis readily allows overcoming the
difficulties within the existing models for cuprates.  The
closed-shell Cu$^{1+}$ ions are now electronically inactive (exactly
as the closed-shell O$^{2-}$ ions are irrelevant in the undoped
limit), and the hole density on the O-ions is nearly $\frac{1}{2}$ in
both electron- and hole-doped cuprates following the transition (since
there are twice as many O-ions as Cu-ions). {\it Period 4 PEC of the
  holes on the O-ions is now a real possibility.} In
Fig.~\ref{figcuprate} we have shown a schematic of the PEC proposed by
us for the O-sublattice with hole charge-density $\frac{1}{2}$.  We
note that: (a) the spin-paired CO lacks C$_4$ symmetry, thus
explaining the simultaneous disappearance of translational and
rotational symmetry (b) the PEC consists of criss-crossed stripes with
periodicities (4a$_0$, 0) and (0. 4a$_0$), as deduced from
experiments.  Computations are currently in progress to investigate
whether or not such a paired CO occurs in the weakly doped
Cu$^{1+}$O$^{2-}$O$^{1-}$ lattice.  These studies will be followed by
search for the PEL state on the same lattice.

\section{Conclusions and Conjectures}

Spin-singlet formation is an essential step to reaching SC. The
emergence of SC from the PEC can therefore be construed as a very
natural phenomenon. In dimensionality greater than 1, however, the
density wave must be commensurate.  Carrier density at or near
$\rho=\frac{1}{2}$ leads naturally to commensurate PEC, which we have
shown from accurate 2D correlated-electron calculations. Computations
further indicate that only at or close to this density
pair-correlations are enhanced by Hubbard $U$, yet another essential
requirement for SC.  Equally importantly, (a) SC in organic CTS is
limited to $\rho=\frac{1}{2}$, even as nonsuperconductors with other
carrier densities exist, and (b) the CO state proximate to SC, when
they both occur, have been found to be the PEC. The idea that SC or
PEL from the $\rho=\frac{1}{2}$ PEC in the CTS thus appears to have a
firm basis.

In the context of the cuprates the valence transition hypothesis, even
as it is orthogonal to existing theories of cuprates, provides a
strong starting point based on which the entire gamut of apparently
contradictory observations can perhaps be explained. It is also
relevant that that there exists no alternate theoretical model
currently that has either found long-range superconducting
correlations, or has demonstrated a Cooper-pair density wave (we
exclude here theoretical models that start from unrealistic
parametrizations; for e.g., models which would give a spin-gapped
state in the undoped limit rather than the experimentally observed
commensurate AFM).  We end this review by pointing out that very
recent experimental works have claimed that even the strange metal
phase in the cuprates evolves from the paired CO phase, and the
charge carriers in this phase are Bosons
\cite{Seibold21a,Yang22a}. Within our theory, the strange metal is a
PEL.  Observation of the strange metal state \cite{Doiron-Leyraud09a}
proximate to the 2k$_F$ CDW-SDW state in (TMTSF)$_2$PF$_6$, that we
have shown to be a PEC (see Section IV.A), supports this viewpoint.

\begin{acknowledgments}
It is a pleasure to contribute this review paper to the special issue
of Chaos dedicated to the 80th birthday of Professor David
K. Campbell. David has been a mentor to and longtime collaborator of
both of us. Mazumdar did his postdoctoral research under David's
supervision at the Center for Nonlinear Studies, Los Alamos National
Laboratory, where he was introduced to the fascinating topic of
nonlinear excitations in organic conjugated polymers. Clay did his
Ph.D. research under David's supervision in the Department of Physics,
University of Illinois at Urbana-Champaign. Some of our earliest works
on the paired-electron crystal, the topic of the present paper, were
results of extended discussions and collaborations with David, that
have continued over the years. We look forward to continued
interactions with David on this and related topics. Earlier research reported
here was supported in part by the NSF and U. S. Department of
Energy. Recent work by SM is supported by NSF-DMR-2301372. Some of the calculations presented here were performed using
the High Performance Computing resources supported by the University
of Arizona and at Mississippi State University (MSU) in the MSU High
Performance Computing Collaboratory (HPC$^2$).
\end{acknowledgments}

\end{document}